\documentclass[12pt]{article}
\usepackage{epsfig}

\begin{document}
\def\br{\begin{eqnarray}}
\def\er{\end{eqnarray}}
\def\be{\begin{equation}}
\def\ee{\end{equation}}
\def\({\left(}
\def\){\right)}
\def\a{\alpha}
\def\b{\beta}
\def\d{\delta}
\def\D{\Delta}
\def\g{\gamma}
\def\G{\Gamma}
\def\h{ {1\over 2}  }
\def\hp{ {+{1\over 2}}  }
\def\hm{ {-{1\over 2}}  }
\def\k{\kappa}
\def\l{\lambda}
\def\L{\Lambda}
\def\m{\mu}
\def\n{\nu}
\def\o{\over}
\def\O{\Omega}
\def\p{\phi}
\def\rh{\rho}
\def\s{\sigma}
\def\t{\tau}
\def\th{\theta}
\def\ii {\'\i  }

\title{{\small {\bf ROTATIONAL CORRECTION ON THE MORSE POTENTIAL THROUGH THE PEKERIS APPROXIMATION AND NIKIFOROV-UVAROV METHOD}}}

\author{C\"uneyt Berkdemir $\thanks{{\small E-mail: berkdemir@erciyes.edu.tr} (C. Berkdemir)}$\\
{\small {\sl Department of Physics, Faculty of Science and
Literature,
Erciyes University,}} {\small {\sl 38039, Kayseri, Turkey}} \\
{\small {\sl and}} \\
{\small {\sl The Abdus Salam International Center for Theoretical
Physics,
Trieste, Italy}}\\\\
Jiaguang Han $\thanks{{\small E-mail: hanjiaguang@sinap.ac.cn} (J. Han)}$ \\
{\small {\sl Shanghai Institute of Applied Physics, Chinese Academy of Science, 201800, Shanghai, China}} \\
{\small {\sl and}} \\
{\small {\sl The Abdus Salam International Center for Theoretical
Physics, Trieste, Italy}}}
\date{}

\maketitle

\begin{abstract}
The Nikiforov-Uvarov method is employed to calculate the the
Schr\"{o}dinger equation with a rotation Morse potential. The
bound state energy eigenvalues and the corresponding eigenfunction
are obtained. All of these calculation present an effective and
clear method under a Pekeris approximation to solve a rotation
Morse model. Meanwhile the results got here are in a good
agreement with ones before.\\
\end{abstract}

\baselineskip=22pt plus 1pt minus 1pt

\vspace{-0.5cm}


\noindent ICTP Serial No.IC/IR/2005/2


\vspace{0.6cm} \baselineskip=18pt
\begin{center}
MIRAMARE -- TRIESTE\\
\smallskip

February 2005\\
\end{center}
\vspace{1.0cm}

\newpage

\maketitle

\section{\small Introduction}
As an empirical potential, the Morse potential has been one of
most useful and convenient model. The Morse potential gives a
excellent qualitative description of the interaction between the
two atoms in a diatomic molecule and also it is a reasonable
qualitative description of the interaction close to the surface
\cite {morse}. For this potential, the Schr\"{o}dinger equation
shall be solved for the angular momentum quantum number $l$ is
equal to zero. As we know the rotation energy of a molecule is
much smaller than that of vibration, and therefore in a pure Morse
potential model the rotation energy of a molecule has been
omitted. However, in some case, it is needed to be included if one
wants to obtain analytical or semianalytical solutions to the
Schr\"{o}dinger equation. Then we will get a rotation Morse
potential. Our interest is how to get the solution of
Schr\"{o}dinger equation for the rotation Morse potential.

In the previous reports, several approximations have been
developed to find better analytical formulas for the rotating
Morse potential \cite{aed}. Some of them require the calculation
of a state-dependent internuclear distance through the numerical
solutions of transcendental equations. Recently, an alternative
method known as the Nikiforov-Uvarov (NU) method is introduced for
solving the Schr\"{o}dinger equation. The application of the
method is to solve Schr\"{o}dinger equation with some well-known
potentials and to solve Dirac, Klein-Gordon and
Duffin-Kemmer-Petiau equation for a Coulomb type potential. In the
present work, the energy and the corresponding eigenfunctions are
calculated employing the NU method under a Pekeris approximation
\cite {pekeris}.

\section{\small Basic Equations of Nikiforov-Uvarov Method}

The NU method provides us an exact solution of non$-$relativistic
Schr\"odinger equation for certain kind of potentials \cite {24}.
The method is based on the solutions of general second order
linear differential equation with special orthogonal functions
\cite {27}. For a given real or complex potential, the
Schr\"odinger equation in one dimension is reduced to a
generalized equation of hypergeometric type with an appropriate
$~s=s(x)~$ coordinate transformation. Thus it can be written in
the following form,
\begin{equation}
\label{eq1} \psi ^{\prime \prime }(s)+\frac{\stackrel{\sim }{\tau
}(s)}{\sigma }\psi ^{\prime }(s)+\frac{\stackrel{\sim }{\sigma
}(s)}{\sigma ^{2}(s)}\psi (s)=0
\end{equation}
where $~\sigma (s)$ and $~\stackrel{\sim }{\sigma }(s)~$ are
polynomials, at most second$-$degree, and $~\stackrel{\sim }{\tau
}(s)~$ is a first$-$degree polynomial. To find a particular
solution of Eq.(\ref{eq1}) by separation of variables, we use the
following the transformation
\begin{equation}
\label{eq2} \psi(s)=\phi(s)y(s)
\end{equation}
This reduces Schr\"odinger equation, Eq.(1), to an equation of
hypergeometric type,
\begin{equation}
\label{eq3} \sigma(s)y^{\prime \prime }+\tau(s)y^{\prime }+\lambda
y=0,
\end{equation}
where $\phi(s)$ satisfies $~\phi (s)^{\prime }/\phi (s)=\pi
(s)/\sigma (s)$. $y(s)~$ is the hypergeometric type function whose
polynomial solutions are given by Rodrigues relation
\begin{equation}
\label{eq4}
y_{n}(s)=\frac{B_{n}}{\rho (s)}\frac{d^{n}}{ds^{n}}\left[ \sigma ^{n}(s)\rho (s)%
\right] ,
\end{equation}
where $B_{n}$ is a normalizing constant and the weight function
$\rho $ must satisfy the condition \cite {24}
\begin{equation}
\label{eq5} (\sigma \rho )^{\prime }=\tau \rho.
\end{equation}
The function$~\pi~$ and the parameter$~\lambda~$ required for this
method are defined as
\begin{equation}
\label{eq6} \pi =\frac{\sigma ^{\prime }-\stackrel{\sim }{\tau
}}{2}\pm
\sqrt{\left(\frac{\sigma^{\prime}-\stackrel{\sim}{\tau}}{2}\right)^{2}-\stackrel{\sim}{\sigma}+k{\sigma}}
\end{equation}
and
\begin{equation}
\label{eq7} \lambda =k+\pi ^{\prime }.
\end{equation}
Here, $\pi(s)$ is a polynomial with the parameter $s$ and the
determination of $k$ is the essential point in the calculation of
$\pi(s)$. Thus, in order to find the value of $k$, the expression
under the square root must be square of a polynomial. Hence, a new
eigenvalue equation for the Schr\"{o}dinger equation becomes
\begin{equation}
\label{eq8} \lambda =\lambda _{n}=-n\tau ^{\prime
}-\frac{n(n-1)}{2}\sigma ^{\prime \prime },~~~(n=0,1,2,...)
\end{equation}
where
\begin{equation}
\label{eq9} \tau (s)=\stackrel{\sim }{\tau }(s)+2\pi (s),
\end{equation}
and it will have a negative derivative.

\section{\small Rotational Correction on the Morse Potential}
Consider a diatomic molecule system with reduced mass $\mu$ in the
Morse potential \be \label{10}
V(r)=D[e^{-2a(r-r_0)}-2e^{-a(r-r_0)}]~~~~~~ (D> 0, a> 0), \ee
where $D$ is dissociation energy, $r_0$ is the equilibrium bond
length and $a$ is a parameter controlling the width of the
potential well. The vibrations of a two-atomic molecule can be
excellently described by this potential type and solved bound
states for $l=0$. If one want to obtain the solution of $l\neq 0$,
the centrifugal term has to be attached to the potential in the
Schr\"odinger radial equation. In this case, total potential shape
becomes as follows \be \label{11}
V_{total}(r)=D[e^{-2a(r-r_0)}-2e^{-a(r-r_0)}]+\frac{\hbar^2l(l+1)}{2\mu
r^2},\ee defining as the sum of the Morse potential and the
centrifugal barrier. In order to calculate the energy eigenvalues
and the corresponding eigenfunction, the potential function given
by Eq.(\ref{10}) is inserted into the radial Schr\"odinger
equation \be \label{12}
\left(-\frac{\hbar^2}{2\mu}\frac{d^2}{dr^2}+D[e^{-2a(r-r_0)}-2e^{-a(r-r_0)}]+\frac{\hbar^2l(l+1)}{2\mu
r^2}\right)R_{nl}(r)=E_{nl}R_{nl}(r). \ee An analytical exact
solution of this differential equation cannot be found without an
approximation. For this case, we now outline the procedure of
Pekeris \cite{pekeris, flugge}.

\subsection{\small Overview of the Pekeris Approximation}

The approximation is based on the expansion of the centrifugal
barrier in a series of exponentials depending on the internuclear
distance, until the second order. However, by construction, this
approximation is valid only for lower vibrational energy states.
Therefore, for a Pekeris approximation, we can take care of the
rotational term in the following way. We let $x=(r-r_0)/r_0$ and
around $x=0$ it may be expanded into a series of powers as \be
\label{13}
V_{rot}(x)=\frac{\gamma}{(1+x)^2}=\gamma(1-2x+3x^2-4x^3+...), \ee
with \be \label{14} \gamma
=\frac{\hbar^2}{2\mu}\frac{l(l+1)}{r_0^2}, \ee the first few terms
should be quite sufficient. Instead, we now replace the rotational
term by the potential \be \label{15} {\tilde{V}_{rot}}(x)= \gamma
\left(D_0+D_1e^{-\alpha x}+D_2e^{-2\alpha x}\right), \ee where
$\alpha = ar_0$. Combining equal powers of Eqs.(\ref{13}) and
(\ref{15}) we get \be \label{16} {\tilde{V}_{rot}}(x)= \gamma
\left(D_0+D_1(1-\alpha
x+\frac{\alpha^2x^2}{2!}-\frac{\alpha^3x^3}{3!}+...)+D_2(1-2\alpha
x+\frac{4\alpha^2x^2}{2!}-\frac{8\alpha^3x^3}{3!}+...)\right), \ee

\be \label{17} {\tilde{V}_{rot}}(x)= \gamma
\left(D_0+D_1+D_2-x(D_1\alpha +
2D_2\alpha)+x^2(D_1\frac{\alpha^2}{2}+2D_2\alpha^2)-x^3(D_1\frac{\alpha^3}{6}+D_2\frac{4\alpha^3}{3})+...\right),
\ee where \br \label {18} D_0=
1-\frac{3}{\alpha}+\frac{3}{\alpha^2}\nonumber \\
D_1=\frac{4}{\alpha}-\frac{6}{\alpha^2}\nonumber \\
D_2=-\frac{1}{\alpha}+\frac{3}{\alpha^2}. \er We now can take the
potential ${\tilde{V}_{rot}}$ instead of the true rotational
potential ${V}_{rot}$ and solve the Schr\"odinger equation for
$l\neq 0$ in Eq.(\ref{12}).

\subsection{\small Applying Nikiforov-Uvarov Method to the Schr\"odinger Equation}

Now, in order to apply the NU$-$method, we rewrite Eq.(\ref{12})
by using a new variable of the form $s=e^{-\alpha x}$,
\begin{equation}
\label{19}
\frac{d^2R_{nl}(s)}{ds^2}+\frac{1}{s}\frac{dR_{nl}(s)}{ds}+\frac{2\mu
r_0^2}{\hbar^2\alpha^2s^2}\left[(E_{nl}-\gamma D_0)+(2D-\gamma
D_1)s-(D+\gamma D_2)s^2\right]R_{nl}(s)=0.
\end{equation}
By introducing the following dimensional parameters
\begin{equation}
\label{20} -\varepsilon_1^2=\frac{2\mu r_0^2(E_{nl}-\gamma
D_0)}{\hbar^2\alpha^2},~~~~~~\varepsilon_2=\frac{2\mu
r_0^2(2D-\gamma
D_1)}{\hbar^2\alpha^2},~~~~~~\varepsilon_3=\frac{2\mu
r_0^2(D+\gamma D_2)}{\hbar^2\alpha^2},
\end{equation}
which leads to a hypergeometric type equation defined in
Eq.(\ref{eq1}):
\begin{equation}
\label{21}
\frac{d^2R_{nl}}{ds^2}+\frac{1}{s}\frac{dR_{nl}}{ds}+\frac{1}{s^2}\times\left[-\varepsilon_1^2+
\varepsilon_2s-\varepsilon_3s^2\right]R_{nl}=0.
\end{equation}
After the comparison of Eq.(\ref{21}) with Eq.(\ref{eq1}), we
obtain the corresponding polynomials as
\begin{equation}
\label{22} \stackrel{\sim}{\tau }(s)=1,~~~{\sigma
}(s)=s,~~~\stackrel{\sim}{\sigma }(s)=-\varepsilon_1^2+
\varepsilon_2s-\varepsilon_3s^2.
\end{equation}
Substituting these polynomials into Eq.(\ref{eq6}), we obtain
$\pi$ function as
\begin{equation}
\label{23} \pi (s)=\pm
\sqrt{\varepsilon_3s^2+(k-\varepsilon_2)s+\varepsilon_1^2}
\end{equation}
taking $ \sigma ^{\prime }(s)=1$. The discriminant of the upper
expression under the square root has to be zero. Hence, the
expression becomes the square of a polynomial of first degree;
\begin{equation}
\label{24} (k-\varepsilon_2)^2-4\varepsilon_1^2 \varepsilon_3=0.
\end{equation}

When the required arrangements are done with respect to the
constant $k$, its double roots are derived as
$k_{+,-}=\varepsilon_2 \pm 2\varepsilon_1 \sqrt{\varepsilon_3}$.

Substituting $k_{+,-}$~into Eq.(\ref{23}), the following possible
solutions are obtained for $\pi (s)$

\begin{equation}
\label{25} \pi(s) = \pm \left\{\begin{array}{ccc}
\left(\sqrt{\varepsilon_3}s-\varepsilon_1~\right), & \hskip 0.5cm
\mbox{for} \hskip 0.5 cm k_-=\varepsilon_2-2\varepsilon_1 \sqrt{\varepsilon_3}\\ \\
\left(\sqrt{\varepsilon_3}s+\varepsilon_1~\right), & \hskip 0.5cm
\mbox{for} \hskip 0.5 cm k_+=\varepsilon_2+2\varepsilon_1 \sqrt{\varepsilon_3}\\
\end{array}\right.
\end{equation}
It is clearly seen that the energy eigenvalues are found with a
comparison of Eq.(\ref{eq7}) and Eq.(\ref{eq8}). From the four
possible forms of the polynomial $\pi (s)$ we select the one for
which the function $\tau (s)$ in Eq.(\ref{eq9}) has a negative
derivative. Therefore, the function $\tau (s)$ satisfies these
requirements, with
\begin{eqnarray}
\label{26}
\tau(s)=1+2\varepsilon_1-2\sqrt{\varepsilon_3}~s, \nonumber \\
\tau^{\prime}(s)=-2\sqrt{\varepsilon_3}~.
\end{eqnarray}
From  Eq.(\ref{eq8}) we also get
\begin{equation}
\label{27} \lambda=\varepsilon_2-2\varepsilon_1 \sqrt{
\varepsilon_3}-\sqrt{\varepsilon_3}~,
\end{equation}
and also
\begin{equation}
\label{28} \lambda=\lambda_n=2n\sqrt{\varepsilon_3}~.
\end{equation}
It is seen that the parameter $-\varepsilon_1^2$ has the following
form
\begin{equation}
\label{29} -\varepsilon _1^2 =-\left[ \frac{\varepsilon
_2}{2\sqrt{\varepsilon _3}}-\left(n+\frac{1}{2}\right) \right]^2.
\end{equation}
Substituting the values of $-\varepsilon_1^2$,  $\varepsilon_2$
and $\varepsilon_3$ into Eq.(\ref{29}), we can immediately
determine the energy eigenvalues $E_{nl}$ as
\begin{equation}
\label{30} E_{nl} =\frac{\hbar^2l(l+1)}{2\mu
r_0^2}\left(1-\frac{3}{ar_0}+\frac{3}{a^2r_0^2}\right)
-\frac{\hbar^2a^2}{2\mu}\left[\frac{\varepsilon
_2}{2\sqrt{\varepsilon _3}}-\left(n+\frac{1}{2}\right)\right]^2,
\end{equation}
where \be \label{31} \frac{\varepsilon _2}{2\sqrt{\varepsilon
_3}}=\frac{1}{a^2\sqrt{\varepsilon _3}}\left[\frac{2\mu
D}{\hbar^2}-\frac{l(l+1)}{r_0^2}\left(\frac{2}{ar_0}-\frac{3}{a^2r_0^2}\right)\right].
\ee \\
The last equation indicates that we deal with a family of the
rotating Morse potential. Of course, it is clear that by imposing
appropriate changes in the parameters $D$, $a$ and $r_0$, the
rotational energy spectrum for the any molecules can be calculated
by the Pekeris approximation and Nikiforov-Uvarov method as well
as other methods \cite {dong, chen, morales, han}. In this study,
we calculate the rotating energy values of the Morse potential for
the CO and LiH molecules. The explicit values of the energy for
the different values of $n$ and $l$ are shown in Tables 1 and 2,
for known values of their relevant potential parameters \cite
{bag, varshni, elso}.

Let us now find the corresponding eigenfunctions for this
potential. Due to the NU$-$method, the polynomial solutions of the
hypergeometric function $y(s)$ depend on the determination of the
weight function $\rho(s)$ which is satisfies the differential
equation $[\sigma (s)\rho (s) ]^{\prime }=\tau (s)\rho (s)$. Thus,
$\rho(s)$ is calculated as
\begin{equation}
\label{32} \rho(s)=s^{1+2\varepsilon
_1}e^{-2\sqrt{\varepsilon_3}~s}.
\end{equation}
Substituting into the Rodrigues relation given in Eq.(\ref{eq4}),
the eigenfunctions are obtained in the following form
\begin{equation}
\label{33} y_{nl}(s)=B_{nl}s^{-(1+2\varepsilon
_1)}e^{2\sqrt{\varepsilon_3}~s}\frac{d^{n}}{ds^{n}}\left[s^{(n+1+2\varepsilon
_1)}e^{-2\sqrt{\varepsilon_3}~s}\right],
\end{equation}
where $B_{nl}$ is the normalization constant. The polynomial
solutions of $y_{nl}(s)$ in Eq.(\ref{33}) are expressed in terms
of the associated Laguerre Polynomials, which is one of the
orthogonal polynomials, that is \be \label{34} y_{nl}(s)\equiv
L_n^{1+2\varepsilon _1}(\nu), \ee where $\nu
=2\sqrt{\varepsilon_3}~s$. By substituting $\pi(s)$ and
$\sigma(s)$ into the expression $\phi (s)^{\prime }/\phi (s)=\pi
(s)/\sigma (s)$ and solving the result differential equation, the
other part of the wave function in Eq.(\ref{eq2}) is found as
\begin{equation}
\label{35} \phi(s)=s^{\varepsilon _1}e^{-\sqrt{\varepsilon_3}~s},
\end{equation}
or in terms of $\nu$
\begin{equation}
\label{36} \phi(\nu)=(2\sqrt{\varepsilon_3})^{-\varepsilon
_1}\nu^{\varepsilon _1}e^{-\nu /2}.
\end{equation}
Combining the Laguerre polynomials and $\phi(\nu)$ in
Eq.(\ref{eq2}), the radial wave functions are constructed as
\begin{equation}
\label{37} R_{nl}(r)=A_{nl}(2\sqrt{\varepsilon_3}~)^{-\varepsilon
_1}\nu^{\varepsilon _1}e^{-\nu /2}L_n^{1+2\varepsilon _1}(\nu),
\end{equation}
where $A_{nl}$ is a new normalization constant. It is clearly
verified that $R_{nl}(r)$ satisfies the following requirement \br
\int_0^\infty R^2_{nl}(r)dr<\infty. \nonumber \er The constant
$A_{nl}$ is determined by making this integral equal 1, \be
\label{38} A_{nl}^2 \frac{(2\sqrt{\varepsilon_3}~)^{-2\varepsilon
_1}}{a}\int_0^\infty \nu^{2\varepsilon
_1-1}e^{-\nu}\left[L_n^{1+2\varepsilon _1}(\nu)\right]^2d\nu=1.
\ee The integral in Eq.(\ref{38}) can be evaluated by using the
recursion relation for Laguerre polynomials and then the
normalization constant can be found as \be \label{39}
A_{nl}^2=\frac{4an!(1+n+\varepsilon_1)^2(2\sqrt{\varepsilon_3}~)^{2\varepsilon
_1}}{(1+n+2\varepsilon_1)!}. \ee Therefore, the simplest radial
wave function becomes for $n=0$: \be \label{40} R_{0l}(r)=
2^{2-\varepsilon_1}(1+\varepsilon_1)\sqrt{\frac{an!}{(1+2\varepsilon_1)!}}~\nu^{\varepsilon_1}~e^{-\nu/2}.
\ee
\section{\small Conclusions}
We have presented an approximation for the rotational correction
on the Morse potential which leads to analytic calculations for
the energy spectrums and wave functions. In these calculations, we
have used a new method which is developed by Nikiforov-Uvarov and
applied the Pekeris approximation. Our main results are summarized
in Eq.(\ref{30}) and Eq.(\ref{37}). This new method is tested by
calculating the energies of some actual rotational states of the
CO and LiH molecules and comparing the results with those of
variational and shifted 1/N expansion methods. The method used
here is the best advantage than the other methods, such as series
solutions, shifted 1/N expansion, supersymmetric approach and
Laplace transforms, due to the fact that a systematical one.\\

\noindent {\bf Acknowledgements}

The authors would like to thank the hospitality and financial
support of the Abdus Salam International Center for Theoretical
Physics, Trieste, Italy, where this work started on February,
2005.

\newpage

{\bf Table 1:}~~{\small Energy eigenvalues (in eV) for the different values of $n$ and $l$ for CO molecule, with $D=90540~cm^{-1}$, $a=2.2994~\AA^{-1}$, $r_0=1.1283~\AA$ and $\mu = 6.8606719~amu$} \cite {bag, elso}.\\

\begin{tabular}{cccc}\hline\hline
\\ $\mathbf {n~~~~~~\ell}
\hspace*{0.2cm} $ & $ \hspace*{0.2cm} \mathbf{NU~Method}
\hspace*{0.2cm} $ & $ \hspace*{0.2cm} \mathbf{Variational}
\hspace*{0.2cm} $ & $ \hspace*{0.2cm} \mathbf{1/N ~Expansion} $ \\[0.3cm]\hline
~~~~&~~~~&~~~~&~~\\[0.2cm]
{0}~~~~~~{0}&~~-11.091~~~&~~-11.093~~&~~~-11.091~~ \\[0.2cm]
{0}~~~~~~{5}&~~-11.084~~&~~-11.085~~&~~~-11.084~~ \\[0.2cm]
{ 0}~~~~~~{10}&~~-11.065~~&~~-11.066~~&~~~-11.065~~ \\[0.2cm]\hline
~~~~&~~~~&~~~~&~~ \\[0.2cm]
{5}~~~~~~{0}&~~-9.795~~&~~-----~~&~~-9.788~~~ \\[0.2cm]
{5}~~~~~~{5}&~~-9.788~~&~~-----~~&~~-9.782~~~ \\[0.2cm]
{ 5}~~~~~~{10}&~~-9.769~~&~~-----~~&~~-9.765~~~ \\[0.2cm]\hline
~~~~&~~~~&~~~~&~~ \\[0.2cm]
{7}~~~~~~{0}&~~-9.299~~&~~-----~~&~~-9.286~~~ \\[0.2cm]
{7}~~~~~~{5}&~~-9.292~~&~~-----~~&~~-9.281~~~ \\[0.2cm]
{ 7}~~~~~~{10}&~~-9.274~~&~~-----~~&~~-9.265~~~ \\[0.2cm]\hline
\end{tabular}

\newpage

{\bf Table 2:}~~{\small Energy eigenvalues (in eV) for the different values of $n$ and $l$ for LiH molecule, with $D=20287~cm^{-1}$, $a=1.1280~\AA^{-1}$, $r_0=1.5956~\AA$ and $\mu = 0.8801221~amu$} \cite {bag, elso}.\\

\begin{tabular}{cccc}\hline\hline
\\ $\mathbf {n~~~~~~\ell}
\hspace*{0.2cm} $ & $ \hspace*{0.2cm} \mathbf{NU~Method}
\hspace*{0.2cm} $ & $ \hspace*{0.2cm} \mathbf{Variational}
\hspace*{0.2cm} $ & $ \hspace*{0.2cm} \mathbf{1/N ~Expansion} $ \\[0.3cm]\hline
~~~~&~~~~&~~~~&~~ \\[0.2cm]
{0}~~~~~~{0}&~~-2.4287~~&~~-2.4291~~~&~~-2.4278 \\[0.2cm]
{0}~~~~~~{5}&~~-2.4012~~&~~-2.4014~~~&~~-2.3999 \\[0.2cm]
{ 0}~~~~~~{10}&~~-2.3287~~&~~-2.3287~~~&~~-2.3261 \\[0.2cm]\hline
~~~~&~~~~&~~~~&~~ \\[0.2cm]
{5}~~~~~~{0}&~~-1.6476~~&~~-----~~~&~~-1.6242 \\[0.2cm]
{5}~~~~~~{5}&~~-1.6236~~&~~-----~~~&~~-1.6074 \\[0.2cm]
{ 5}~~~~~~{10}&~~-1.5606~~&~~-----~~~&~~-1.5479 \\[0.2cm]\hline
~~~~&~~~~&~~~~&~~ \\[0.2cm]
{7}~~~~~~{0}&~~-1.3774~~&~~-----~~~&~~-1.3424 \\[0.2cm]
{7}~~~~~~{5}&~~-1.3549~~&~~-----~~~&~~-1.3309 \\[0.2cm]
{ 7}~~~~~~{10}&~~-1.2957~~&~-----~~&~~-1.2781 \\[0.2cm]\hline
\end{tabular}


\newpage

\begin{thebibliography}{99}
\bibitem{morse} P. M. Morse, Pyhs. Rev. {\bf 34}, 57 (1929).
\bibitem{aed} A. E. DePristo, J. Chem. Phys. {\bf 74}, 5037 (1981).
\bibitem{pekeris} C. L. Pekeris, Pyhs. Rev. {\bf 45}, 98 (1934).
\bibitem{24}
A. F. Nikiforov, V. B. Uvarov, "Special Functions of Mathematical
Physics" (Birkhauser, Basel, 1988).
\bibitem{27} G. Szego, "Orthogonal Polynomials", (American Mathematical Society, New York, Revised edition, 1959).
\bibitem{flugge} S. Fl\"ugge, "Practical Quantum Mechanics I" (Springer-Verlang, Berlin,
1971).
\bibitem{dong} S. H. Dong, and G. H. Sun, Physics Letters {\bf A
314}, 261 (2003).
\bibitem{chen} G. Chen, Physics Letters {\bf A 326}, 55 (2004).
\bibitem{morales} D. A. Morales, Chemical Physics Letters {\bf
394}, 68 (2004).
\bibitem{han} D. Han, X. Song, and X. Yang, Physica {\bf A 345}, 485 (2005).
\bibitem{bag} M. Bag, M. M. Panja, R. Dutt and Y. P. Varshni, Pyhs.
Rev. {\bf A 46}, 6059 (1992).
\bibitem{varshni} Y. P. Varshni, Can. J. Chem. {\bf 66}, 763 (1988).
\bibitem{elso} E. D. Filho, Phys. Lett. {\bf A 269}, 269 (2000).
\end{thebibliography}
\end{document}